\documentstyle[aps,epsf]{revtex}
\renewcommand{\thefootnote}{\alph{footnote}}

\title{Excitation Function of K$^+$ and $\pi^+$
Production in Au+Au Reactions at 2-10 AGeV
\footnote{Corresponding Author: C.A. Ogilvie, 
24-410 Massachusetts Institute of Technology, Cambridge MA 02139,
ogilvie@mit.edu}}

\begin{document}
\input epsf
\maketitle
\renewcommand{\thefootnote}{\alph{footnote}}
\begin{center}
\it{E866 Collaboration}
\end{center}

L. Ahle$^{10}$\footnote{Lawrence Livermore National Laboratory, Livermore CA 94550}, 
Y.~Akiba$^{6}$, K.~Ashktorab$^{2}$, 
M.D.~Baker$^{10}$\footnote{Brookhaven National Laboratory, Upton, NY 11973}, D. Beavis$^{2}$,
B.~Budick$^{11}$, J.~Chang$^{3}$, C.~Chasman$^{2}$, Z.~Chen$^{2}$, 
Y.Y.~Chu$^{2}$, T.~Chujo$^{15}$, 
J.~Cumming$^{2}$, R.~Debbe$^{2}$, J.C.~Dunlop$^{10}$\footnote{Yale University, New Haven, CT 06520},  
W.~Eldredge$^{3}$,
K.~Fleming$^{3}$,
S.-Y.~Fung$^{3}$, E.~Garcia$^{9}$, S.~Gushue$^{2}$,  H.~Hamagaki$^{14}$,
R.~Hayano$^{13}$,  G.H.~Heintzelman$^{10b}$, 
J.H.~Kang$^{16}$, E.J.~Kim$^{2,16}$,  A.~Kumagai$^{15}$,
K.~Kurita$^{15}$, J.H.~Lee$^{2}$, Y.K.~Lee$^{8}$, Y.~Miake$^{15}$, 
A.C.~Mignerey$^{9}$, 
M.~Moulson$^{4}$\footnote{Laboratori Nazionali di Frascati, INFN, 00044 Frascati, Italy}, 
C.~Muentz$^{2}$\footnote{Goethe Universit\"at, Institut f\"ur Kernphysik, Frankfurt, Germany}, 
K.~Nagano$^{13}$, C.A.~Ogilvie$^{10}$, J.~Olness$^{2}$,  K.~Oyama$^{13}$, 
L.~Remsberg$^{2}$,  H.~Sako$^{15}$, 
R.~Seto$^{3}$,  J.~Shea$^{9}$,  K.~Shigaki$^{7}$, S.G.~Steadman$^{10}$, 
G.S.F.~Stephans$^{10}$,  T.~Tamagawa$^{13}$,
M.~Tannenbaum$^{2}$,  S.~Ueno-Hayashi$^{15}$,
F.~Videbaek$^{2}$, H.~Xiang$^{3}$, F.~Wang$^{4}$\footnote{Lawrence Berkeley National
Laboratory, Berkeley, CA 94720},  K.~Yagi$^{15}$, 
F.~Zhu$^{2}$,\\

\begin{center}
\it{E917 Collaboration}
\end{center}

B.B.~Back$^{1}$, R.R.~Betts$^{1,5}$, J.~Chang$^{3}$,
W.C.~Chang$^{3}$\footnote{Institute of Physics, Academica Sinica, Taipei 11529, Taiwan}, 
C.Y.~Chi$^{4}$, Y.Y.~Chu$^{2}$, J.B.~Cumming$^{2}$,
J.C.~Dunlop$^{10c}$, W.~Eldredge$^{3}$, S.Y.~Fung$^{3}$, 
R.~Ganz$^{5}$\footnote{Max-Planck-Institut f\"ur Physik, D-80805 M\"unchen, Germany},
E.~Garcia$^{9}$, A.~Gillitzer$^{1}$\footnote{Institut f\"ur Kernphysik, 
Forschungzentrum J\"ulich,D-52425 J\"ulich, Germany}, G.H.~Heintzelman$^{10b}$,
W.F.~Henning$^{1}$, D.J.~Hofman$^{1}$, B.~Holzman$^{5}$, J.H.~Kang$^{16}$,
E.J.~Kim$^{2,16}$, S.Y.~Kim$^{16}$, Y.~Kwon$^{16}$,
D.~McLeod$^{5}$, A.C.~Mignerey$^{9}$, M.~Moulson$^{4,d}$,
V.~Nanal$^{1}$\footnote{Tata Institute of Fundamental Research, Colaba, Mumbai 400005 India}, 
C.A.~Ogilvie$^{10}$, R.~Pak$^{12}$,
A.~Ruangma$^{9}$, D.~Russ$^{9}$, R.~Seto$^{3}$, P.J.~Stanskas$^{9}$,
G.S.F.~Stephans$^{10}$, H.~Wang$^{3}$, F.L.H.~Wolfs$^{12}$, 
A.H.~Wuosmaa$^{1}$,
H.~Xiang$^{3}$, G.H.~Xu$^{3}$, H.B.~Yao$^{10}$, C.M.~Zou$^{3}$
\\
\begin{center}
{\it
$^{1}$ Argonne National Laboratory, Argonne, IL 60439
\\
$^{2}$ Brookhaven National Laboratory, Upton, NY 11973
\\
$^{3}$ University of California Riverside, Riverside, CA 92521
\\
$^{4}$ Columbia University, Nevis Laboratories, Irvington, NY 10533
\\
$^{5}$ University of Illinois at Chicago, Chicago, IL 60607
\\
$^{6}$ High Energy Accelerator Research Organization 
(KEK), Tanashi-branch, (Tanashi) Tokyo 188, Japan\\
$^{7}$ High Energy Accelerator Research Organization 
(KEK),Tsukuba, Ibaraki 305, Japan\\
$^{8}$ Johns Hopkins University, Baltimore, MD 
\\
$^{9}$ University of Maryland, College Park, MD 20742
\\
$^{10}$ Massachusetts Institute of Technology, Cambridge, MA 02139
\\
$^{11}$ New York University, New York, NY 
\\
$^{12}$ University of Rochester, Rochester, NY 14627
\\
$^{13}$ Department of Physics, University of Tokyo, Tokyo 113, Japan 
\\
$^{14}$ Center for Nuclear Study, School of Science, University of Tokyo, 
Tanashi, Tokyo 188, Japan\\
$^{15}$ University of Tsukuba, Tsukuba, Ibaraki 305, Japan \\
$^{16}$ Yonsei University, Seoul 120-749, South Korea
\\
}
\end{center}
\begin{abstract}
Positive pion and kaon production
from Au+Au reactions have been measured as a function of
beam energy over the range 2.0-10.7~AGeV.
Both the kaon and the pion production cross-sections
at mid-rapidity are observed to increase
steadily with beam kinetic energy. The
ratio of K$^+$ to $\pi^+$ mid-rapidity yields
increases from 0.0271$\pm0.0015\pm0.0014$ at 2.0~AGeV to 
0.202$\pm0.005\pm0.010$            
at 10.7~AGeV and is larger than the  K$^+$/$\pi^+$ ratio
from p+p reactions over the same beam energy region.
There is no indication
of an onset of any new production mechanism 
in heavy-ion reactions in this energy range beyond rescattering
of hadrons.  
\end{abstract}


\vspace*{0.5cm}

This Letter presents the results of the first measurements
of particle production in Au+Au reactions 
in the energy range from 2 to 10~AGeV.
Recent results\cite{Chen98} for proton stopping in high
energy nuclear collisions
suggest that an extended region of
dense nuclear
matter is formed at beam energies 
near 10~AGeV.
Increasing the beam energy from 2~A to 10~AGeV is expected
to increase the maximum density achieved from
approximately twice normal density at 
2~AGeV\cite{Sto86,Ber88,Cas90,Bau91} to an
estimated eight times normal nuclear density at 
10~AGeV\cite{Li95}. 

The properties of such dense matter can be
explored by measuring how the production of pions and kaons varies
with beam energy.  
For example, secondary collisions may occur between 
resonanant states, and the exciation energy of the resonances is
then available for particle production.
At a beam energy of 10.7~AGeV
the kaon yield per participant
in central Au+Au reactions\cite{kpkm}
has been measured to be three to four times that for
p+p reactions at the same energy, consistent with the majority of
kaon production coming from secondary rather than primary
hadronic collisions.
At lower beam energies, a larger fraction 
of these secondary collisions will be below
the threshold for producing a pair of strange hadrons.
Yet, in models of nucleus-nucleus reactions at 1-2~AGeV\cite{Li95,kaos},
kaon production is dominated by
the rare secondary collisions that
are above threshold compared to production from 
primary collisions boosted by Fermi motion. 
An excitation function of heavy-ion reactions from 1 to 10~AGeV can 
therefore be
used to study how the
role of secondary collisions in strangeness production
evolves with increasing beam energy.

Secondary collisions can also change the 
production of pions.
At beam energies near 1~AGeV, the measured yield
of pions per participant
in central Au+Au reactions is below the yield of pions in nucleon-nucleon
collisions at the same energy\cite{Pel97,Mue97}.
Transport calculations of heavy-ion reactions at these beam 
energies\cite{Bau91} reproduce this reduction in pion yield 
through the
production of $\Delta$'s and their subsequent rescattering
$\Delta+N\rightarrow N+N$. 
At higher energies the effects of this rescattering may be
reduced, since
it has been observed that the mid-rapidity $\pi$ yields in p+A
reactions at 14.6AGeV\cite{pA} are independent of A. 
In addition, the measured pion yield per participant from
central Si+A reactions at 14.6~AGeV/c\cite{SiA} is similar to that measured in 
peripheral reactions, in contrast to the reduction observed in Au+Au
reactions at 1~AGeV\cite{Pel97,Mue97}.
The transverse spectra of pions 
may also provide information on the reaction dynamics.
Spectra from Au+Au collisions at 10.7~AGeV have been observed to
rise faster than an exponential function at low p$_T$\cite{Chen98}, with 
$\pi^-$ having a larger low-p$_T$ rise than $\pi^+$.
There have been many 
suggestions for the origin of this rise: 
pion emission from different stages of the reaction\cite{Bauer},
the decay of resonances\cite{res}, and possible 
enhancement due to Bose-statistics\cite{Bose}. 
To make progress on these topics, it is useful to measure
the evolution of the mid-rapidity pion spectra and yields as a function
of beam energy from 1 to 10~AGeV. 

It is also possible that a small region of baryon-rich quark-gluon plasma 
might be formed in heavy-ion reactions near 10AGeV. This could be observable as
changes in the characteristics of pion and kaon production
as the beam energy increases.
 
The data in this Letter come from two separate AGS collaborations, E866 and 
E917; each of which used much the
same apparatus.  The E866 collaboration measured
Au+Au collisions at 1.96, 4.00, and 10.7~AGeV 
kinetic energy. 
The E917 collaboration measured Au+Au reactions at 5.93 and 7.94~AGeV 
kinetic energy. The beam energies quoted and used in this Letter correspond 
to the energy half-way through the target.
For shorthand these will be referred to as 2, 4, 6, 8 and 10.7 AGeV
in the text.
The pion and kaon spectra at 10.7~AGeV have already been 
published by E866\cite{Chen98,kpkm}. 

The new data presented in this
Letter (Au+Au at 2, 4, 6, and 8~AGeV)
were measured with the Henry Higgins 
spectrometer used by experiments E802, E859, E866 and E917.
For more details on the equipment and data analysis the
reader is referred to references\cite{Chen98,kpkm,nim}; a brief description is
provided here.
A Au target of 1960 mg/cm$^2$ was used for each data set. This thickness 
corresponds to approximately a 3\% interaction probability.
The Au beam loses approximately 0.06 AGeV as it passes through the target.
The rotatable  spectrometer has an
acceptance of 25~msr with an opening angle of 
$\Delta\theta\sim 14^\circ$, and consists of drift and multi-wire chambers 
on either side of a dipole magnet. 
Similar track reconstruction, efficiency corrections, acceptance and decay
algorithms were used for the data at each
beam energy.
Particle identification was performed using
a wall of time-of-flight detectors with a timing resolution, $\sigma$, of 130~ps.  
The detectors for global event characterization were
a multiplicity array surrounding the target, and a calorimeter placed 
at zero degrees with a 1.5$^\circ$ opening angle.
For the 2, 4, 6, and 8~AGeV Au+Au data, 
central collisions were selected with the multiplicity
array.
The data at 10.7~AGeV  were measured prior to installation of 
the multiplicity detector, so 
central collisions at this beam energy were selected using
the zero-degree calorimeter.
A comparison of event selection using these two devices has been 
published\cite{double}. 
Based on similar analyses fully described in 
references\cite{Chen98,kpkm}
the total systematic uncertainty in the  
normalization of the particle yields is 15\% and is 
dominated by the uncertainty in the acceptance, tracking efficiency, 
and extrapolation to low transverse 
mass  required to obtain particle yields.
The systematic uncertainty on the mean transverse mass
extracted from the spectra is estimated to be 10\%.
The relative beam-energy-to-beam-energy
systematic uncertainty in both the yields and mean transverse mass
is 5\% because,
while the systematic uncertainty from the  
geometrical acceptance is common to all the measurements,
the dependence of the tracking efficiency on the occupancy of the
spectrometer\cite{kpkm} and the uncertainty 
involved in extrapolating to obtain particle yields are not.

\begin{figure}[htb]
\begin{minipage}[t]{75mm}
\epsfxsize=7.5cm\epsfbox[10 90 450 500]{
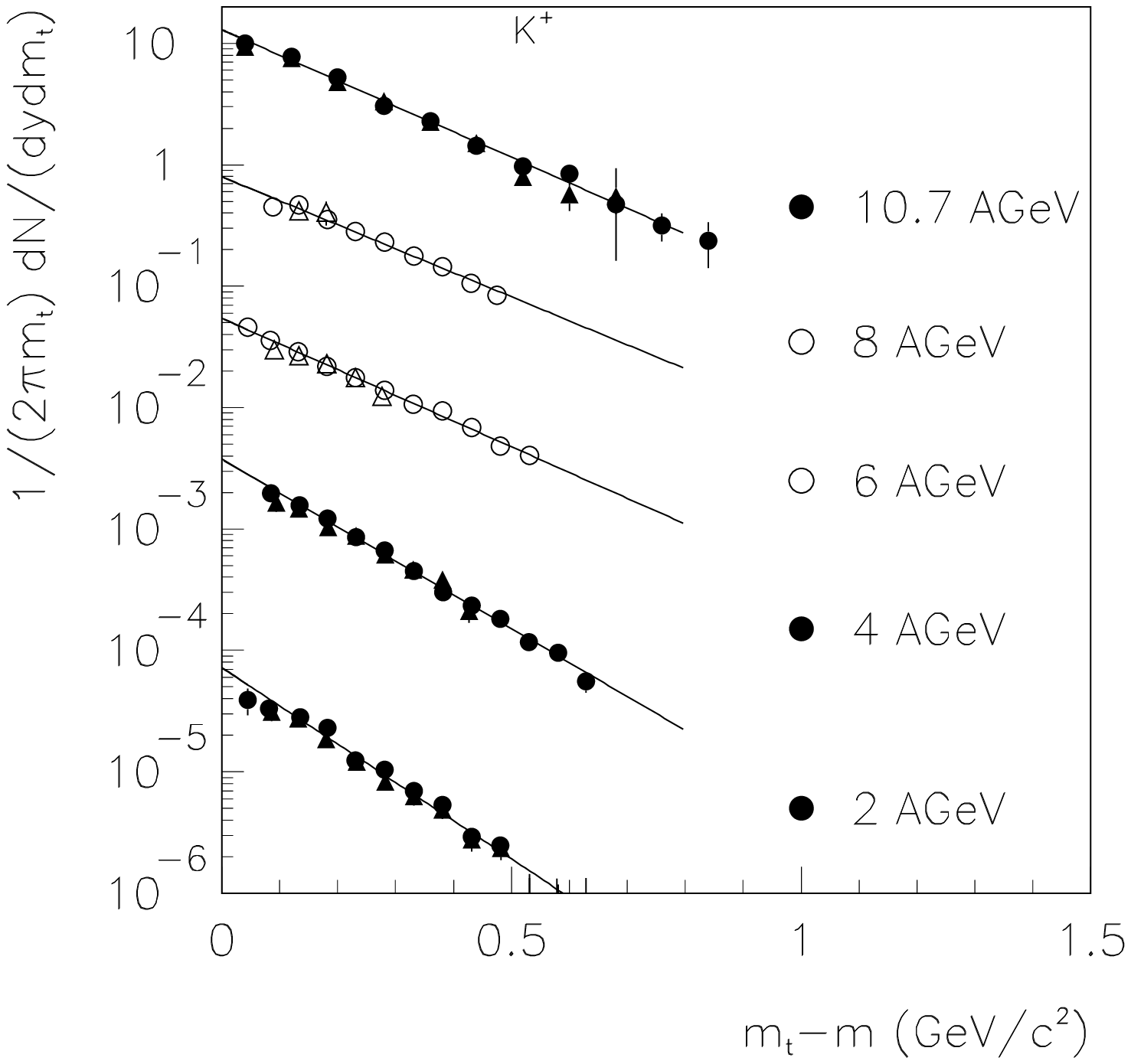}
\end{minipage}
\hspace{\fill}
\begin{minipage}[t]{85mm}
\epsfxsize=7.5cm\epsfbox[10 90 450 500]{
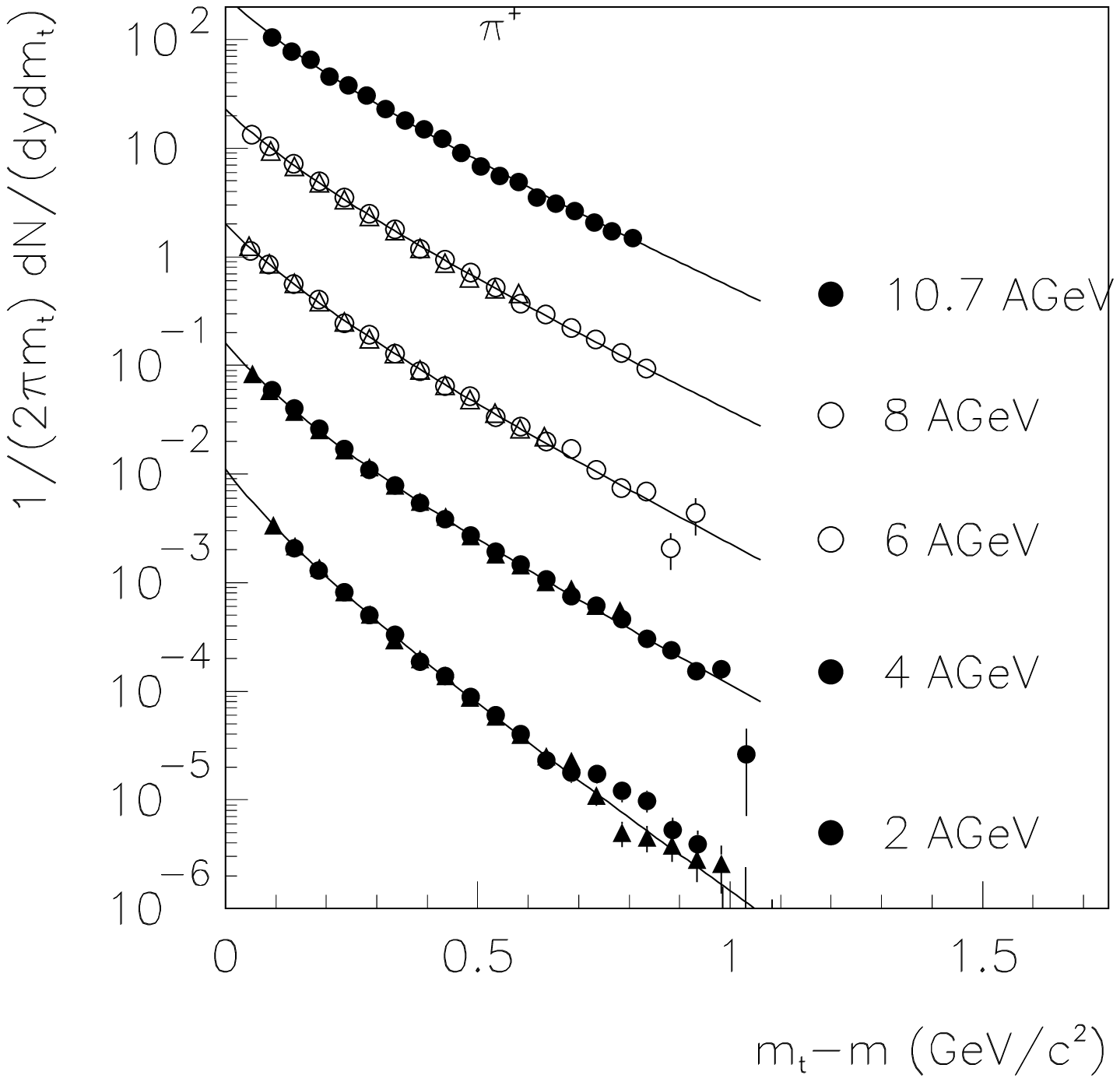}
\end{minipage}
\caption{The invariant yield of K$^+$ (left panel) and $\pi^+$
 (right panel) as a function 
of m$_t$ at mid-rapidity from Au+Au collisions
for the different beam energies.
For each energy the spectrum just back of mid-rapidity is shown with 
circles,
the spectrum just forward of mid-rapidity is shown with triangles.   
The data from E866 are shown as filled symbols and the data from
E917 are shown as open symbols. The data at 10.7~AGeV are shown at the
correct scale, the data at each lower energy are 
divided by successive powers of ten for clarity.
The errors are statistical only.
}
\label{fig:mtkaon}
\end{figure}

The invariant yields of K$^+$ are 
shown in the left panel of Figure \ref{fig:mtkaon}   
as a function of transverse 
mass, m$_t=\sqrt{p_t^2+m_0^2}$, for
Au+Au central collisions at 2, 4, 6, 8, 
and 10.7~AGeV\cite{kpkm}  beam kinetic 
energy.
The centrality selection at each beam energy is the most central
5\% of the total interaction cross-section ($\sigma_{int}$=6.8b).
For each beam energy there are 
two m$_t$ spectra shown in Figure
\ref{fig:mtkaon}.
In each of the  2, 4, 6, and 8~AGeV data sets,  
one of these spectra corresponds to a 
rapidity slice just backwards of mid-rapidity,  
$-0.25<\frac{y-y_{nn}}{y_{nn}}<0$ (shown as circles in 
Figure \ref{fig:mtkaon}) and  the other is symmetrically 
forward of mid-rapidity,
$0<\frac{y-y_{nn}}{y_{nn}}<0.25$ 
(shown as triangles), where $y_{nn}$ is mid-rapidity in the laboratory frame.
For the 10.7~AGeV reactions 
the rapidity slice is narrower,
$|\frac{y-y_{nn}}{y_{nn}}|<0.125$  to match the bin-width of the
published pion spectra\cite{Chen98}.  
A single exponential function in m$_{t}$ was 
fit simultaneously to the two kaon spectra at each beam energy 
\begin{equation}
\frac{1}{2\pi m_t}\frac{d^2N}{dm_tdy}=
\frac{dN/dy}{2\pi(Tm_0 + T^2)}e^{-(m_t-m_0)/T} 
\end{equation}
The fits reproduce the spectra well with two free parameters,
the inverse slope
parameter T and the rapidity density, dN/dy, in that
rapidity slice. These parameters
are tabulated in Table~\ref{tabkaon} for each beam
energy, as well as the mean value of the transverse mass, 
$\langle m_t \rangle$, calculated from the inverse slope parameter.
The kaon spectra were  also fit by the scaled exponential
given in Equation 2 below. For all energies except 8.0~AGeV, 
these fits produced dN/dy and $\langle m_t \rangle$ values within 
5\% of the values extracted from the exponential fits. The yields from the 
scaled exponential fits at 8.0~AGeV were 10\% smaller than the exponential
fits. 
This is consistent with the estimated systematic uncertainty.
 
The right panel of Figure~\ref{fig:mtkaon} 
shows invariant spectra for $\pi^+$ from 
Au+Au reactions at each of the five beam energies.
The 10.7~AGeV data have already been published\cite{Chen98}.
The spectra cover the same rapidity ranges as for the kaons.
These spectra 
rise above an exponential at low m$_t$, and were therefore fit
with a scaled exponential,
\begin{equation}
\frac{1}{2\pi m_t}\frac{d^2N}{dm_tdy}=\frac{dN/dy}
{2\pi T_s^{2-\lambda}\Gamma(2-\lambda,m_0/T_s)}m_t^{-\lambda}
e^{-m_t/T_s}~~~~~.
\label{scalel}
\end{equation}              
$\Gamma(2-\lambda,m_0/T_s)$, the complementary incomplete gamma function,
is introduced in the normalization
so that dN/dy is a fitted parameter.
The yields and the other free parameters, $\lambda$ and 
T$_s$, are tabulated in Table~\ref{tabpion}
along with $\langle m_t \rangle$ calculated from the fit
parameters in Equation~\ref{scalel}
\begin{equation}
<m_t>=\frac{T_s\Gamma(3-\lambda,m_0/T_s)}{\Gamma(2-\lambda,m_0/T_s)}.
\end{equation}
The uncertainty in $\langle m_t \rangle$ for pions includes the
covariance between the fitted parameters $\lambda$ and T$_s$.

The pion invariant yields could also be fit with
a sum of two exponentials in m$_t$. 
These fits produced dN/dy and $\langle m_t \rangle$ values which are within
5\% of the values obtained from the scaled exponential fits.
Scaled exponential
fits to the pion transverse spectra are used
throughout this paper.

\begin{figure}[htb]
\epsfysize=12cm\epsfbox[1 60 450 505]{
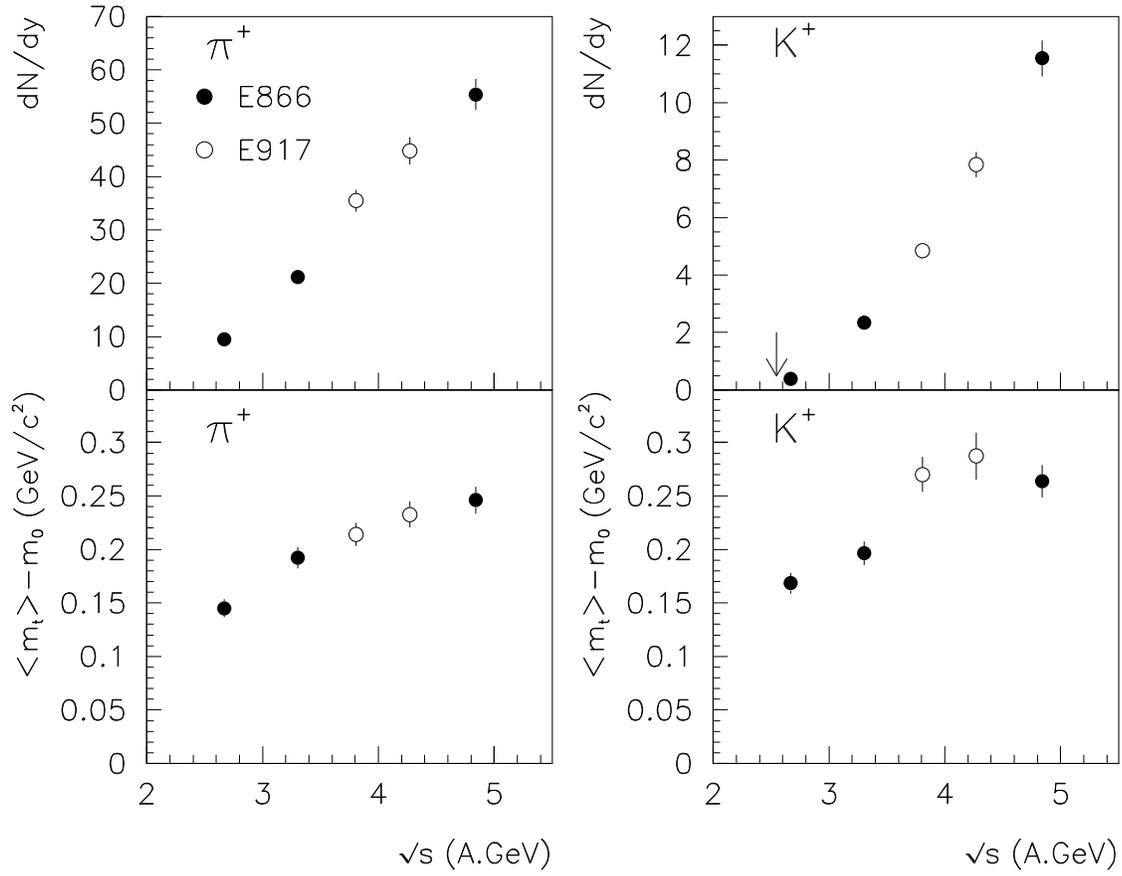}
\caption{The yield of $\pi^+$ and K$^+$ at mid-rapidity (top-panels)
for central Au+Au
reactions as a function of the initial available beam energy.
The data from E866 are shown as filled circles and the data from
E917 are shown as open circles. The lower panels
show the mean m$_t$ minus the rest mass
for $\pi^+$ and K$^+$ at 
the same rapidity.
The errors include both statistical and a 5\% (dN/dy), 5\% (mean m$_t$)
point-to-point systematic uncertainty. 
The arrow indicates the threshold energy for producing K$^+$ in a 
p+p reaction. 
}
\label{fig:excit}
\end{figure}

The mid-rapidity yields of $\pi^+$ and 
K$^+$ are shown in the upper panels
of Figure~\ref{fig:excit} plotted as a function of 
the c.m. energy of initial nucleon-nucleon collisions, $\sqrt{s}$.  
Both  the $\pi$
and the K yields increase steadily with beam energy.   
In the lower panels of Figure~\ref{fig:excit}, 
the values of $\langle m_t \rangle$ minus the rest mass ($m_0$) for
$\pi^+$ and K$^+$ are plotted versus $\sqrt{s}$.
Compared to the increase 
of particle production, $\langle m_t \rangle$ is observed to increase
more slowly with beam energy.                                     
Within the statistical and systematic errors  
the changes in $\langle m_t \rangle$-$m_0$ for pions and kaons
with beam energy are similar.
It is noted that the $\lambda$ parameter 
from the scaled exponential fits to pions (equation \ref{scalel}
and Table \ref{tabpion}) slightly decreases as the beam 
energy increases, i.e. the $m_t$ spectrum is tending towards a pure exponential with
increasing beam energy.                            

\begin{figure}[htb]
\epsfxsize=12cm\epsfbox[30 60 520 465]{
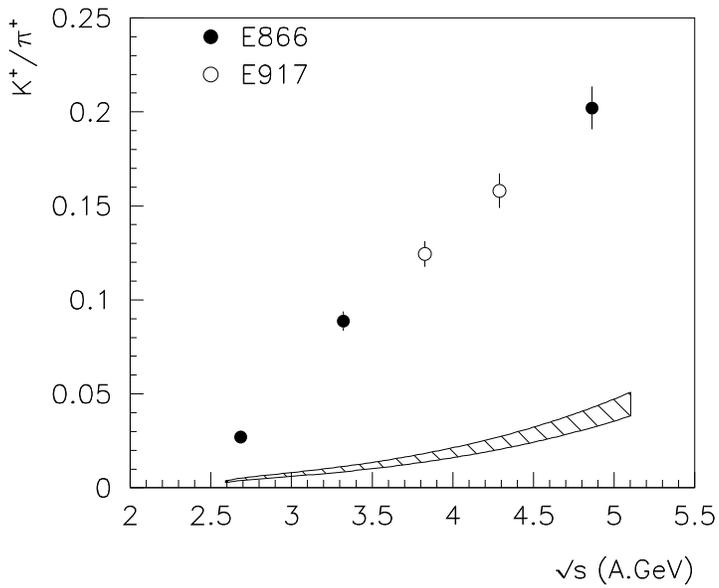}
\caption{The ratio of dN/dy for K$^+$/$\pi^+$ at mid-rapidity in
central Au+Au reactions as a function 
of the initial available energy.
The data from E866 are shown as filled circles and the data from
E917 are shown as open circles.
The errors include a 5\% systematic uncertainty.
The hashed region is the K$^+$/$\pi^+$ ratio from the parameterized K and 
$\pi$ yields from p+p reactions (see text for details). The hashed region
covers $\pm1\sigma$ around the p+p K$^+$/$\pi^+$ ratio.
}
\label{fig:kpi}
\end{figure}

The increase in K yield  
with beam energy shown in Figure 2 
is more rapid than the increase of $\pi$ yield.  
This is emphasized in 
Figure~\ref{fig:kpi}, 
where the ratio of dN/dy for $K^+$ to $\pi^+$ is 
plotted versus $\sqrt{s}$.
There is a 5\% point-to-point systematic
uncertainty in this quantity.
The measured $K^+/\pi^+$ ratio increases steadily from 
0.0271$\pm0.0015\pm0.0014$ at 2~AGeV to 
0.202$\pm0.005\pm0.010$ at 
10.7~AGeV. 
The smooth increase of the measured K$^+$/$\pi^+$ ratio
suggests that there
is no onset of a production mechanism 
different than hadronic scattering as the beam energy is 
increased. It is noted that ratios can be difficult to interpret
because changes could occur in both the numerator
and denominator. 
The measured ratio K$^+/\pi^+ = 0.19\pm 0.01$ from 
Pb+Pb collisions at 157~AGeV/c\cite{NA49} is comparable to the ratio
K$^+/\pi^+ = 0.202\pm0.025\pm0.010$ from Au+Au reactions at 10.7~AGeV
of the present work.
This suggests that either the ratio saturates or that a maximum exists in
the K$^+$/$\pi^+$ from heavy-ion reactions at 
energies between the AGS and SPS. 
The mid-rapidity K$^+$/$\pi^+$ ratio
from central Au+Au reactions at 1~AGeV measured by the KaoS
collaboration\cite{kaos} is K$^+$/$\pi^+=(3\pm 1)\times 10^{-3}$.
This is one order of magnitude below
the K$^+$/$\pi^+$ ratio at 2~AGeV from the present work.

Data for inclusive K$^+$ yields in p+p reactions in the same energy
range as the Au data in this Letter (3 to 12~GeV/c)
are available\cite{Fes79,Reed68,lb,Ant}.
A suitable parameterization of these yields
was suggested in reference \cite{Sib}
\begin{equation}
\rm{Y_{K+}}=c\times(s/s_0 - 1)^a \times (s/s_0)^b \hspace*{0.2in} 
\sqrt{s_0}<\sqrt{s}<20 GeV
\label{ks}
\end{equation}
where s$_0=(m_p+m_K+m_\Lambda)^2$. 
The p+p data were fit to
obtain the parameters a=0.223, b=2.196, and c=0.00221.
From the errors of the data points it is estimated that 
there is a 10\% 
systematic uncertainty in the parameterized kaon yield 
from $3<\sqrt{s}<5$ GeV. 
The $\pi^+$ yields from p+p reactions have been fit
by Rossi et al.\cite{Rossi}  with
\begin{equation}
\rm{Y_{\pi+}}=a+b\times ln(s)+c/\sqrt{s} \hspace*{0.2in}
3<\sqrt{s}<20 GeV
\label{pis}
\end{equation} We use the parameters obtained by Rossi et al.\cite{Rossi};
a= -1.55, b=0.82, and c=0.79. 
From the errors on the data points it is estimated that the 
systematic uncertainty of the parameterized pion yield is 10\%.

In this energy range, the K$^+$ yield from p+p reactions 
increases faster with
beam energy than the $\pi^+$ yield, as is shown in
Figure~\ref{fig:kpi} where the
ratio of the p+p K$^+$ and $\pi^+$ yields is shown as a hashed region.
The measured K$^+$/$\pi^+$ ratio in Au+Au reactions 
is significantly larger than the ratio from p+p reactions.
The data from Au+Au are measured at 
mid-rapidity whereas the p+p results are integrated over the full
phase space. 
As an estimate of the level of the difficulties this might cause, 
in Au+Au reactions at 10.7~AGeV the 
mid-rapidity K$^+$/$\pi^+$ ratio is 0.202$\pm0.005\pm0.010$ and 
is within a few percent of the value obtained by integrating over a
broader rapidity range of $0.6<y<2.0$ where 
K$^+$/$\pi^+=0.197\pm0.003\pm0.010$\cite{kpkm}.
Also the production data from p+p reactions are used in this comparison instead
of estimating n+p, n+n and then averaging over the initial isospin.
The isospin-averaged K$^+$/$\pi^+$ ratio for N+N reactions is within 20\% of
the K$^+$/$\pi^+$ ratio for p+p reactions at 10.7 AGeV\cite{Ahle}. 

To summarize, particle production in central Au+Au reactions
has been measured at
2, 4, 6, 8 and 10.7~AGeV beam kinetic energy.
This data set bridges the gap
between experiments at the AGS and BEVALAC/SIS accelerators,
and as such provides a strong test for hadronic models that have been 
developed and applied only at one or the other end of this 
excitation function. 
The mid-rapidity yields of both pions and kaons from central
Au+Au reactions
steadily increase with beam energy.
Over this energy range the mean m$_t$ of the
transverse spectra of pions and kaons increases by less than 75\%.
A larger fraction of the extra 
available energy therefore goes into particle
production rather than into increasing the transverse energy per particle. 
The shape of the pion transverse spectrum has a low m$_t$ rise
above an exponential function. This rise is slightly smaller at
the higher energies. 
The K$^+/\pi^+$ ratio increases steadily with beam
energy from nearly 3\% at 2~AGeV to 20\% at 10.7~AGeV.
The K$^+/\pi^+$ ratio is larger in Au+Au reactions than
in p+p reactions. 
The smooth increase of the K$^+$/$\pi^+$ ratio
in Au+Au suggests that there
is no evidence for an onset of a production mechanism 
other than hadronic scattering as the beam energy is 
increased. 

This work was supported by the Department of Energy (USA), the National
Science Foundation (USA), NASA (USA), Ministry of Education, Science, Sports and 
Culture (Japan), and by KOSEF (Korea)

\begin{table}[hbt]
\begin{tabular}{|c|c|c|c|c|c|} 
\hline
K$^+$ & E$_{kin}$  & y$_{nn}$ & dN/dy & $T$  & $<m_t>$-m$_0$ \\ 
& (AGeV) & & & GeV/c$^2$ &  GeV/c$^2$  \\ \hline
E866 &1.96 & 0.90 & 0.381 $\pm$ 0.015 & 0.138 $\pm$ 0.004  & 0.168 $\pm$ 0.005 \\
E866 &4.00 & 1.17 & 2.34 $\pm$ 0.05 & 0.158 $\pm$ 0.003 &  0.197 $\pm$ 0.005 \\
E917& 5.93 & 1.34 & 4.84 $\pm$ 0.09 & 0.208 $\pm$ 0.006  & 0.270 $\pm$ 0.009 \\
E917& 7.94 & 1.47 & 7.85 $\pm$ 0.21 & 0.219 $\pm$ 0.011 & 0.287 $\pm$ 0.017 \\
E866& 10.74 & 1.60 & 11.55 $\pm$ 0.24 & 0.204 $\pm$ 0.006 & 0.264 $\pm$ 0.008\\
\hline 
\end{tabular}
\parbox{15.0cm}{\caption [] {\label{tabkaon}
Excitation function of K$^+$ spectral characteristics
at mid-rapidity from central Au+Au reactions.                
The mid-rapidity range 
for 2, 4, 6, 8 AGeV is $|\frac{y-y_{nn}}{y_{nn}}|<0.25$,
for 10.7 AGeV the width is  $|\frac{y-y_{nn}}{y_{nn}}|<0.125$,
where $y_{nn}$ is mid-rapidity in the laboratory frame.  
The errors are statistical only.
}}
\end{table}
\clearpage

\begin{table}[hbt]
\begin{tabular}{|c|c|c|c|c|c|} 
\hline
$\pi^+$ & E$_{kin}$  & dN/dy & $T_{s}$ & $\lambda$ & $<m_t>$-m$_0$\\ 
& (AGeV) &  & GeV/c$^2$ & & GeV/c$^2$ \\ \hline
E866 &1.96 & 14.1 $\pm$ 0.5 & 0.149 $\pm$ 0.008 & 1.06 $\pm$ 0.18 & 0.145 $\pm$ 0.004 \\
E866 &4.00 & 26.4 $\pm$ 0.4 & 0.205 $\pm$ 0.009 & 1.14 $\pm$ 0.10 & 0.192 $\pm$ 0.002 \\
E917& 5.93 & 38.9 $\pm$ 0.5 & 0.205 $\pm$ 0.010 & 0.91 $\pm$ 0.11 & 0.214 $\pm$ 0.003 \\
E917& 7.94 & 49.7 $\pm$ 0.7 & 0.215 $\pm$ 0.012 & 0.84 $\pm$ 0.11 & 0.233 $\pm$ 0.003 \\
E866& 10.74 & 57.1 $\pm$ 0.8 & 0.230 $\pm$ 0.008 & 0.87 $\pm$ 0.09 & 0.246 $\pm$ 0.003\\
\hline 
\end{tabular}
\parbox{15.0cm}{\caption [] {\label{tabpion}
Excitation function of $\pi^+$ spectral characteristics
at mid-rapidity from central Au+Au reactions.
The mid-rapidity range 
for 2, 4, 6, and 8 AGeV beam energy is $|\frac{y-y_{nn}}{y_{nn}}|<0.25$,
for 10.7 AGeV is  $|\frac{y-y_{nn}}{y_{nn}}|<0.125$,
where $y_{nn}$ is the mid-rapidity in the laboratory frame.  
The errors are statistical only.
}}
\end{table}
\clearpage

\clearpage
\clearpage

\end{document}